\newcommand\apjcls{1}
\newcommand\aastexcls{2}
\newcommand\othercls{3}

\newcommand\papercls{\aastexcls}
\documentclass[tighten, times, trackchanges, preprint2]{aastex6}

\if\papercls \apjcls
\usepackage{apjfonts}
\else\if\papercls \othercls
\usepackage{epsfig}
\fi\fi
\usepackage{ifthen}
\usepackage{natbib}
\usepackage{amssymb, amsmath}
\usepackage{mathrsfs}
\usepackage{bm}
\usepackage{appendix}
\usepackage{etoolbox}
\usepackage[T1]{fontenc}
\usepackage{paralist}

\if\papercls \apjcls
\newcommand\aas{\ref@jnl{AAS Meeting Abstracts}}
\newcommand\dps{\ref@jnl{AAS/DPS Meeting Abstracts}}
\newcommand\maps{\ref@jnl{MAPS}}
\else\if\papercls \othercls
\usepackage{astjnlabbrev-jh}
\fi\fi

\if\papercls \aastexcls
\hypersetup{citecolor=blue, 
            linkcolor=blue, 
            menucolor=blue, 
            urlcolor=blue}  
\else
\usepackage[
bookmarks=true,           
bookmarksnumbered=true,   
colorlinks=true,          
citecolor=blue,           
linkcolor=blue,           
menucolor=blue,           
urlcolor=blue,            
linkbordercolor={0 0 1},  
pdfborder={0 0 1},
frenchlinks=true]{hyperref}
\fi

\if\papercls \othercls

\else

\fi

\providecommand{\adsurl}[1]{\href{#1}{ADS}}

\makeatletter
\patchcmd{\NAT@citex}
  {\@citea\NAT@hyper@{%
     \NAT@nmfmt{\NAT@nm}%
     \hyper@natlinkbreak{\NAT@aysep\NAT@spacechar}{\@citeb\@extra@b@citeb}%
     \NAT@date}}
  {\@citea\NAT@nmfmt{\NAT@nm}%
   \NAT@aysep\NAT@spacechar\NAT@hyper@{\NAT@date}}{}{}

\patchcmd{\NAT@citex}
  {\@citea\NAT@hyper@{%
     \NAT@nmfmt{\NAT@nm}%
     \hyper@natlinkbreak{\NAT@spacechar\NAT@@open\if*#1*\else#1\NAT@spacechar\fi}%
       {\@citeb\@extra@b@citeb}%
     \NAT@date}}
  {\@citea\NAT@nmfmt{\NAT@nm}%
   \NAT@spacechar\NAT@@open\if*#1*\else#1\NAT@spacechar\fi\NAT@hyper@{\NAT@date}}
  {}{}
\makeatother

\makeatletter
\DeclareRobustCommand{\lowcase}[1]{\@lowcase#1\@nil}
\def\@lowcase#1\@nil{\if\relax#1\relax\else\MakeLowercase{#1}\fi}
\makeatother

\DeclareSymbolFont{UPM}{U}{eur}{m}{n}
\DeclareMathSymbol{\umu}{0}{UPM}{"16}
\let\oldumu=\umu
\renewcommand\umu{\ifmmode\oldumu\else\math{\oldumu}\fi}

\if\papercls \othercls

\else

\fi

\let\oldsim=\sim
\renewcommand\sim{\ifmmode\oldsim\else\math{\oldsim}\fi}
\let\oldpm=\pm
\renewcommand\pm{\ifmmode\oldpm\else\math{\oldpm}\fi}
\newcommand\by{\ifmmode\times\else\math{\times}\fi}

\newcommand\tablebox[1]{\begin{tabular}[t]{@{}l@{}}#1\end{tabular}}
\newbox{\wdbox}
\renewcommand\c{\setbox\wdbox=\hbox{,}\hspace{\wd\wdbox}}
\renewcommand\i{\setbox\wdbox=\hbox{i}\hspace{\wd\wdbox}}

\newcount\timect
\newcount\hourct
\newcount\minct
\newcommand\now{\timect=\time \divide\timect by 60
         \hourct=\timect \multiply\hourct by 60
         \minct=\time \advance\minct by -\hourct
         \number\timect:\ifnum \minct < 10 0\fi\number\minct}

\catcode`@=11

\newcommand\comment[1]{}

\newcommand\commenton{\catcode`\%=14}

\renewcommand\math[1]{$#1$}
\newcommand\mathshifton{\catcode`\$=3}

\let\atab=&
\newcommand\atabon{\catcode`\&=4}

\let\oldmsp=\sp
\let\oldmsb=\sb
\def\sp#1{\ifmmode
           \oldmsp{#1}%
         \else\strut\raise.85ex\hbox{\scriptsize #1}\fi}
\def\sb#1{\ifmmode
           \oldmsb{#1}%
         \else\strut\raise-.54ex\hbox{\scriptsize #1}\fi}
\newbox\@sp
\newbox\@sb
\def\sbp#1#2{\ifmmode%
           \oldmsb{#1}\oldmsp{#2}%
         \else
           \setbox\@sb=\hbox{\sb{#1}}%
           \setbox\@sp=\hbox{\sp{#2}}%
           \rlap{\copy\@sb}\copy\@sp
           \ifdim \wd\@sb >\wd\@sp
             \hskip -\wd\@sp \hskip \wd\@sb
           \fi
        \fi}
\def\msp#1{\ifmmode
           \oldmsp{#1}
         \else \math{\oldmsp{#1}}\fi}
\def\msb#1{\ifmmode
           \oldmsb{#1}
         \else \math{\oldmsb{#1}}\fi}

\def\supon{\catcode`\^=7}

\def\subon{\catcode`\_=8}

\def\supsubon{\supon \subon}

\newcommand\actcharon{\catcode`\~=13}

\newcommand\paramon{\catcode`\#=6}

\comment{And now to turn us totally on and off...}

\newcommand\reservedcharson{ \commenton  \mathshifton  \atabon  \supsubon 
                             \actcharon  \paramon}

\catcode`@=12
\reservedcharson

\if\papercls \apjcls

\else

\fi

\if\papercls \othercls
\else
  \newcommand\inpress{n}
  \if\inpress y
    \received{\today}
    \revised{}
    \accepted{}
    \if\papercls \apjcls
    \slugcomment{}
    \fi
  \else
  \slugcomment{\tablebox{Submitted to {\em ApJ}. DRAFT of {\today}.}}
  \fi
\fi

\newcommand\chisq{\ifmmode{\chi\sp{2}}\else\math{\chi\sp{2}}\fi}
\newcommand\redchisq{\ifmmode{ \chi\sp{2}\sb{\rm red}}
                    \else\math{\chi\sp{2}\sb{\rm red}}\fi}
\newcommand\Teq{\ifmmode{T\sb{\rm eq}}\else$T$\sb{eq}\fi}
\newcommand\mjup{\ifmmode{M\sb{\rm Jup}}\else$M$\sb{Jup}\fi}
\newcommand\rjup{\ifmmode{R\sb{\rm Jup}}\else$R$\sb{Jup}\fi}
\newcommand\msun{\ifmmode{M\sb{\odot}}\else$M\sb{\odot}$\fi}
\newcommand\rsun{\ifmmode{R\sb{\odot}}\else$R\sb{\odot}$\fi}
\newcommand\mearth{\ifmmode{M\sb{\oplus}}\else$M\sb{\oplus}$\fi}
\newcommand\rearth{\ifmmode{R\sb{\oplus}}\else$R\sb{\oplus}$\fi}

\shorttitle{Rapidly Rotating Black Holes in HMXRBs}
\shortauthors{Batta, Ramirez-Ruiz \& Fryer}

\begin{document}

\title{The Formation of Rapidly Rotating Black Holes in High Mass X-ray Binaries}

\author{Aldo Batta\altaffilmark{1},
Enrico Ramirez-Ruiz\altaffilmark{1},
and
Chris Fryer\altaffilmark{2,3}
}

\affil{\sp{1} Department of Astronomy and Astrophysics, University of California, Santa Cruz, CA 95064, USA\\
       \sp{2} Department of Physics, The University of Arizona, Tucson, AZ 85721, USA.\\
       \sp{3} CCS Division, Los Alamos National Laboratory, Los Alamos, NM 87545, USA.}
\begin{abstract}
High mass X-ray binaries (HMXRBs)  like Cygnus X-1, host some of the most rapidly spinning black holes (BHs) known to date, reaching spin parameters $a \gtrsim 0.84$. However, there are several effects that can severely limit the maximum BH spin parameter that could be obtained from direct collapse, such as tidal synchronization, magnetic core-envelope coupling and mass loss. Here we propose an alternative  scenario where the BH is produced by  a {\it failed}  supernova (SN) explosion that is unable to  unbind the stellar progenitor. A large amount of fallback material ensues, whose interaction with the secondary  naturally increases its overall angular momentum content, and therefore, the spin of the BH when accreted. Through SPH hydrodynamic simulations, we studied the unsuccessful explosion of a $8M_{\odot }$ pre-SN star in a close binary with a $12M_{\odot}$ companion with an orbital period of $\approx1.2$ days, finding that it is possible to obtain a BH with a high spin parameter $a\gtrsim0.8$ even when the expected  spin parameter from direct collapse is $a \lesssim 0.3$. This scenario also naturally explains the atmospheric metal pollution  observed in  HMXRB stellar companions. 
\end{abstract}



\section{INTRODUCTION}
\label{intro}

To date, we know of the existence of a handful of HMXRBs harboring  stellar mass BHs. The BHs in these systems are characterized by having a large spin parameter $a>0.8$ \citep{Gou2009,Gou2014,Liu2008}, which are difficult to  reconcile with the classical formation scenario of HMXRBs where the BH is formed from the direct collapse (or {\it failed} supernova) of a massive star in a close binary. A HMXRB can result in this scenario provided that there is low mass loss from the {\it failed} supernova (SN) and the BH experiences either a small natal kick that induces no significant orbital change, or a strong kick in a  restricted  direction that results in a final close orbit. For stellar binaries forming short-period HMXRBs  ($\lesssim 1$ day), tidal synchronization  is expected to  be effective \citep{vandenHeuvel2007}. This tidally locking constraints  the total angular momentum content available to form the BH. Even in the absence  tidal locking,  the maximum BH spin parameter can be restricted by the interior of  the  star being spun  down  either  by  magnetic  coupling  \citep{Spruit2002} or mass loss. 

Assuming that synchronization induces  rigid body rotation \citet{Lee2002} calculated the BH spin expected from direct collapse. They found that for typical  orbital periods (of a few days), the maximum spin parameter is $a\lesssim 0.4$.\footnote{For comparison, the spin of a BH expected  from the collapse of a maximally rotating polytropic star is  $a\approx0.75$  \citep{Shapiro2002}.} What is more, the final spin of the BH depends on the mass distribution of the pre-SN  star, which differs significantly from a polytropic one. In fact, more realistic pre-SN  mass distributions yield more compact cores, which produce even smaller spin parameters, thus severely hampering the formation of rapidly rotating BHs in typical HMXRBs.

Alternatively, as has been observed in red giants \citep{Beck2012}, it is possible that the core of the pre-SN star decouples from the slower rotating, tidally locked outer layers. Such rapidly rotating cores, commonly found in some pre-SN progenitor models \citep{Heger2000,Woosley2006}, could produce a rapidly rotating BH from the collapse of the Fe core. However,  when tidal forces are effective in bringing a significant fraction of the stellar progenitor to synchronous rotation, a BH formed from a rapidly rotating core will be naturally slowed down as it accretes the slowly rotating outer layers. This implies that the only way to get a rapidly rotating BH from a star with a slowly rotating envelope is from the interaction with the companion or from external mass accretion. 

The accretion rate in HMXRBs can, however, only be a small fraction of the companion's mass loss rate, typically $\lesssim 10^{-5}M_{\odot}\ \rm yr^{-1}$ \citep{Crowther2007}. This in turn places an upper limit on the amount of mass that the BH could have accreted, given the short lifetime of the companion ($\approx 10^5$ yr for WR stars). While high BH spins in low mass XRBs  can be acquired via mass transfer \citep{Fragos2015}, as shown by \citet{King1999} and \citet{Podsiadlowski2003}, this is not the case for HMXRBs. In such systems it has been widely believed that the properties of BHs are primarily  natal \citep{Liu2008,Valsecchi2010,Axelsson2011}. 

Here we propose an alternative  scenario  in which the spin of the BH comes from accretion of fallback material ejected during a {\it failed} SN explosion. To this end, we ran a series of 3d Smoothed Particle Hydrodynamics (SPH) simulations of  the ejection and fallback of the low velocity, tidally locked stellar envelope in a close binary with a massive companion and an orbital period that is typical of HMXRBs. In what follows, we study the properties of  pre-SN stars and the low-energy SN explosions that  can result in large BH spins  in  HMXRBs as well as  atmospheric metal pollution in  the surviving stellar companions.


\section{BH formation and fallback from failed SN in binaries}

In the classical formation scenario, the HMXRB progenitor begins  as a detached binary with mass ratio  $q=M_{\rm p}/M_{\rm c} > 1$. As the primary of mass $M_{p}$ evolves, a mass transfer episode is triggered. With $q>1$, the mass transfer is likely to become unstable once the donor develops an extended convective envelope or evolves to become a red supergiant. This will shrink the binary orbit so efficiently that a common envelope phase will be triggered \citep{Tauris2006}. As a result, the primary can end up as a He star accompanied by a more massive companion  with an orbital period of a few days. Tidal synchronization of its components might not be attained before the primary reaches the end of its life \citep{vandenHeuvel2007}, however, magnetic coupling between  the core and the stelar envelope \citep{Spruit2002}, or mass loss preceding BH formation could strongly reduce the amount of angular momentum in the primary.

Once the primary exhausts it's nuclear fuel and reaches the pre-supernova  stage, the collapse of the Fe core will lead to the formation of a proto-neutron star (PNS), which will inject energy into the infalling stellar layers through its copious neutrino emission. The shock produced by this  energy injection will move outwards decelerating the infalling layers and forcing  them out. Shocked material with velocities smaller than the escape velocity, will ultimately fallback onto the newly formed PNS. The subsequent formation of a BH will depend on the amount of  fallback material, which in turn, depends strongly on the energy injected and the structure of the star \citep{Fryer2012,Ugliano2012,Pejcha2015a}.  It is believed that  the nature of  {\it failed} SNe might be responsible for  shaping  the galactic black hole mass function \citep{Ozel2010,Kochanek2014a,Kochanek2014b}. 

The outcome of the {\it failed} SNe scenario could range from the quiet disappearance of a star producing no observable transient, to the production of low luminosity  transients \citep{Fryer2009,Moriya2010,Piro2013,Lovegrove2013,Dexter2013}. The amount and structure of fallback material will be mainly determined  by the currently unknown properties  of the explosion, which in turn depends on the debated structure of the stellar progenitor at the end of its life \citep{Perna2014}.  Recent observational evidence suggests that these {\it failed} explosions may not be that uncommon  \citep{Kochanek2008,Reynolds2015,Adams2016} and that the assembly  of BH binaries  (GW150914 and GW151226)  observed by LIGO may necessitate a {\it failed} supernova scenario \citep{Abbott2016, Abbott2016c,Belczynski2016} in order  to  explain the large  BH masses inferred.

A weak or {\it failed} SN explosion is expected to be produced during the collapse of a non-negligible  fraction of massive stars \citep{Nadezhin1980,Lovegrove2013,Piro2013}. In such instances, energy injection  fails to completely unbind  the star such that a  sizable fraction of its mass eventually fallbacks onto the newly formed  BH. If this takes place  within a close binary, a fraction of the stellar material can expand to a radius comparable or larger than the binary's separation, thus engulfing the companion. The capture of such low velocity SN ejecta by a close companion could explain the overabundances of $\alpha$ elements (O, Mg, Si, S) observed in the companion stars of X-ray binary systems \citep{Israelian1999,Podsiadlowski2002,Willems2005,Gonzalez2008, Gonzalez2011, Suarez2015}. Eventually, the interaction with the companion can torque the fallback material around the newly formed BH, thus effectively changing  its angular momentum content.  It is to this issue  that we now turn our attention.


\subsection{Setup for SPH simulations}
To study whether or not the interaction of fallback material with the recently formed BH-star binary could translate into a sizable  increase in the overall angular momentum of the material accreted onto the BH, we perform  a series of 3d SPH  simulations using a modified version of \textsc{gadget2} \citep{Springel2005}. In particular,  we study the expansion  and fallback of a low velocity, tidally locked stellar envelope in a close binary hosting a $12M_{\odot}$ companion in a 1.2 days  orbit. The density, velocity and internal energy profile of the envelope was derived from the SN explosion of an $8M_{\odot}$ helium core of a $25M_{\odot}$ KEPLER model from \cite{Woosley2002}, with slightly varied parameters as described in Fryer et al (2017, in preparation). The collapse is modeled using a 1-dimensional core-collapse code including a coupled equation of state covering a broad range of densities, a gray flux-limited, three-species neutrino diffusion scheme, a small nuclear network and a spherically-symmetric, post-newtonian GR routine \citep{Herant1994,Fryer1999}.  After the collapse and bounce, we remove the core and use a parameterized energy injection routine mimicking the convection-enhanced supernova engine to produce a range of explosion energies and remnant masses.  Our study focuses on an explosion that produces reasonable remnant masses for our study. We considered the innermost 2 solar masses of the pre-SN  star to be the newly formed BH with an initial spin $a=0$, which we treat as a sink particle. 

\begin{figure}[!h]
\begin{center}
\includegraphics[trim={0.65cm 0 1.5cm 0},width=0.48\textwidth]{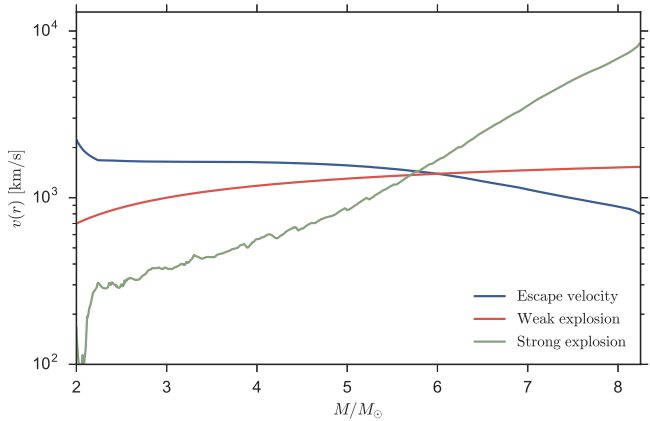}

   \caption{Radial velocity profiles (green and red lines) and escape velocity (blue line) as a function of the inner mass $M(r)$ contained within the radius $r$, for the two explosion scenarios explored here. The (relatively) strong (green line)  and  {\it failed} (red line) SN explosions have kinetic to binding energy ratios of  $\sigma = E_{\rm kin}/E_{\rm bin} = 3.75$ and  $\sigma=0.77$, respectively. Both explosions have very similar fallback masses of $M_{\rm fb}\approx 6M_{\odot}$, which include the initial mass of the BH.}
 \label{fig:Vprof}

 \end{center}
 \end{figure}

Three different scenarios are explored. The first considers the direct collapse of the envelope without any energy injection.  The second uses the successful though weak SN model discussed above, in which the explosion has a kinetic to binding energy ratio $\sigma = E_{\rm kin}/E_{\rm bin} = 3.75$ (where $E_{\rm bin}=1.3\times10^{50} \rm erg$) resulting in the high velocity ejection of about 1/3 of the envelope. In the third scenario, we adopted a shallower velocity profile, representing a very weak ({\it failed}) SN explosion with a kinetic to binding energy ratio $\sigma =  0.77$. Figure \ref{fig:Vprof} shows the two different velocity profiles used in our simulations as a function of the inner mass coordinate $M$. The blue line represents the escape velocity of the envelope, and the green and red lines represent the velocity profiles for the successful  and   {\it failed} SN explosions, respectively. Both explosions have a very similar fallback mass of  $M_{\rm fb}\approx  6M_{\odot}$ but very different velocity profiles, which leads to a very different interaction between the fallback material and the binary. These different  interactions  result in varying  accretion histories, which ultimately determine the BH's final mass and spin.

The 1d SN explosion models are mapped  into a 3d spherically symmetric distribution of SPH particles. This distribution  is obtained by constructing spherical shells with $N_{\rm shell}$ particles evenly distributed across the shell's surface using HEALPix \citep{Gorski2005}. The number of particles in each shell is determined by the local density $\rho(r)$, the total number of SPH particles $N_{\rm sph}$ and the particle masses $m_{i}=M_{t}/N_{\rm sph}$. After running a series of convergence tests for the accretion rate onto the BH with $N_{\rm sph}$ ranging from  $5\times10^5$ to $1\times10^7$, we settled for a resolution of $N_{\rm sph}\simeq 2\times10^6$.

\subsection{Accretion onto the BH}

Since the resolution of the simulation gradually decreases as  particles are accreted and  material escapes, we had to implement an accretion  prescription. An accretion radius of $r_{\rm acc}<0.01R_{\odot}$ from the BH is defined, within which particles are  accreted if certain conditions are met. Particles within $r_{\rm acc}$ falling towards the BH with less specific angular momentum than the one needed to orbit at the innermost circular stable orbit ($j<j_{\rm isco}$) are considered to be accreted. These particles transfer their entire mass and angular momentum to the BH. Particles within $r_{\rm acc}$, falling towards the BH and with specific angular momentum $j_{\rm crit}\leq j < 10 j_{\rm crit}$ are also considered to be accreted. These particles are assumed to be accreted via an accretion disk in a viscous time scale  $t_{\nu}\approx  \alpha^{-1}(H/R)^{-2} 1/\Omega_{\rm k}$,  where  $\Omega_{\rm k}=(GM_{\rm bh}/r_{\rm disk}^3)^{1/2}$,  $H/R\approx 0.1$ and  $\alpha\approx  0.1$. These particles  transfer $j=j_{\rm isco}$ to the BH (with the rest assumed to be effectively  transported outwards) and have a fraction of their rest mass radiated  away as in \citet{Bardeen1970a} and \citet{Thorne1974}.


\section{Results}
\begin{figure}[!h]
\begin{center}
 \begin{minipage}[c]{0.5\textwidth}
\includegraphics[trim={0.45cm 0.2cm 0.9cm 0.7cm},width=1.0\textwidth]{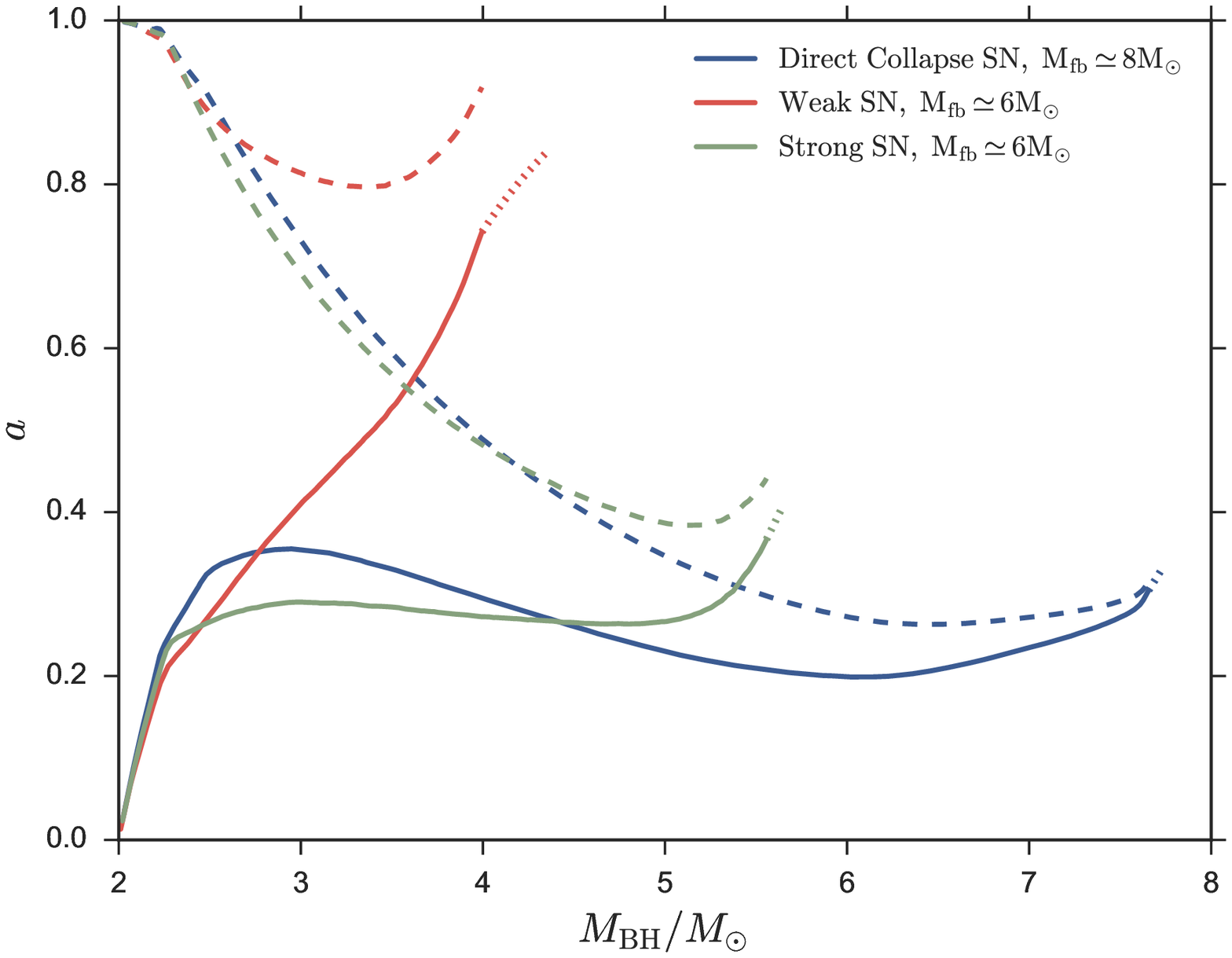}
\includegraphics[trim={0.45cm 0 0.9cm 0.9cm},width=1.0\textwidth]{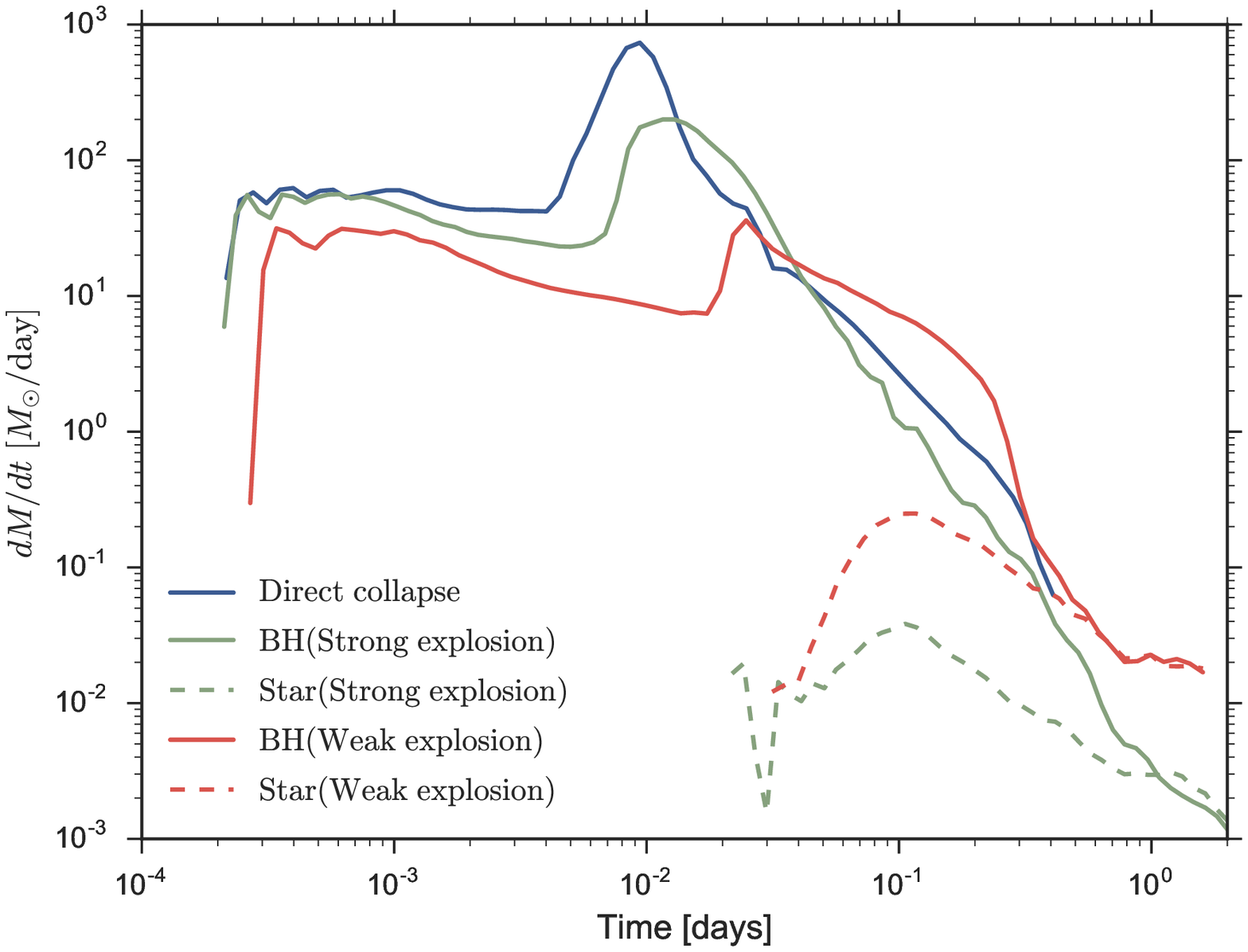}
\end{minipage}%
   \caption{The spin parameter of the BH (top panel) as a function of the accreted mass for the three different scenarios explored here: a  {\it failed} SN explosion  (red line), a successful SN  explosion (green line), and direct collapse  (blue line).  The solid lines correspond to our simulations, and the dashed line indicates the evolution of the BH's spin assuming that a maximally rotating BH was formed from the core and accretes the same angular momentum and mass as in the simulations. The bottom panel shows the accretion rate evolution obtained for the BH and the companion.}
 \label{fig:Bhspin}

 \end{center}
 \end{figure}

Figure \ref{fig:Bhspin} shows the evolution of the BH's spin parameter as a function of the accreted  mass together with the corresponding  accretion rates obtained for the three scenarios explored here. The accretion of subcritical material ($j<j_{\rm crit}$) transfers its angular momentum directly to the BH, while supercritical material ($j\gtrsim j_{\rm crit}$) is only accreted after a viscous timescale, transferring $j= j_{\rm crit}$ \citep{Bardeen1970a,Thorne1974}. Thus, there is a limit in the maximum amount of specific angular momentum that can be transferred to the BH via a disk.  In general, the BH accretes material  with mainly subcritical angular momentum which only marginally increases $a$. 

The evolution of the BH spin in our simulations is shown as solid lines in the top panel of Figure \ref{fig:Bhspin}. The BH has an initial spin of $a\simeq0$ which increases as mass and angular momentum is accreted. In order to consider the scenario where the BH forms from a rapidly rotating core, which is decoupled from the synchronized envelope, we also calculate the evolution of an initially maximally rotating BH ($a\simeq0.99$) accreting material as in our simulations. The evolution of this maximally rotating BH is shown as dashed lines in the top panel of Figure \ref{fig:Bhspin}. Each line is colored according to the three scenarios explored: {\it blue} for direct collapse, {\it green} for a strong SN explosion and {\it red} for a weak SN explosion.  Since the spin of the BH and its mass determine the value of $r_{\rm isco}$ and the angular momentum accreted through a disk $j_{\rm isco}$, the evolution of the maximally rotating BH's is only an approximation, and overestimates to the total amount of angular momentum being accreted through a disk.

As can be seen in the lower panel from Figure \ref{fig:Bhspin}, the accretion rates for the simulations with  successful  and  {\it failed} SN explosions decrease to $\dot{M}\lesssim 10^{-2}M_{\odot}\ \rm day ^{-1}$ after $t\simeq1$ day. With such low accretion rates, the simulation would require a prohibitively large CPU time to modestly  increase the mass of the BH. We  thus decided to stop the simulations at this point and obtain an estimate of the spin parameter assuming that all  of the material  bounded to the BH (and within its Roche lobe) will eventually be accreted through an accretion disk. The transition between the solid and dashed line shown in the top panel of Figure \ref{fig:Bhspin} marks the end of the accretion rates derived from the simulation and beginning of the analytical estimate  \citep{Bardeen1970a,Thorne1974}. This gives a robust upper limit to the spin parameter obtained from the accretion of such material. This estimate does not account for material outside the BH's Roche lobe, which although still  bounded to the system it is  likely to be accreted instead by the more massive companion. 

As expected, the direct collapse simulation gives rise to the smallest spin parameter ($a\approx  0.3$), which is without surprise  similar  to the one obtained by \citet{Lee2002}. In the successful SN explosion calculation, a prompt  collapse of  the innermost $\approx 3.5M_{\odot}$  follows by the ejection  of $\approx 2 M_{\odot}$ of envelope material. In this case,  only about $\approx 0.5M_{\odot}$ of material is  able to be effectively torqued by the binary. In the  {\it failed} SN explosion case, there is a substantial  amount of material that interacts with the binary and, as a result, a circumbinary disk is produced. This results in a significant  increase in the angular momentum content  of  the fallback material, which in turn  increases the BH spin to $a\gtrsim0.8$.

\begin{figure}[!h]
\begin{center}

\includegraphics[width=0.49\textwidth]{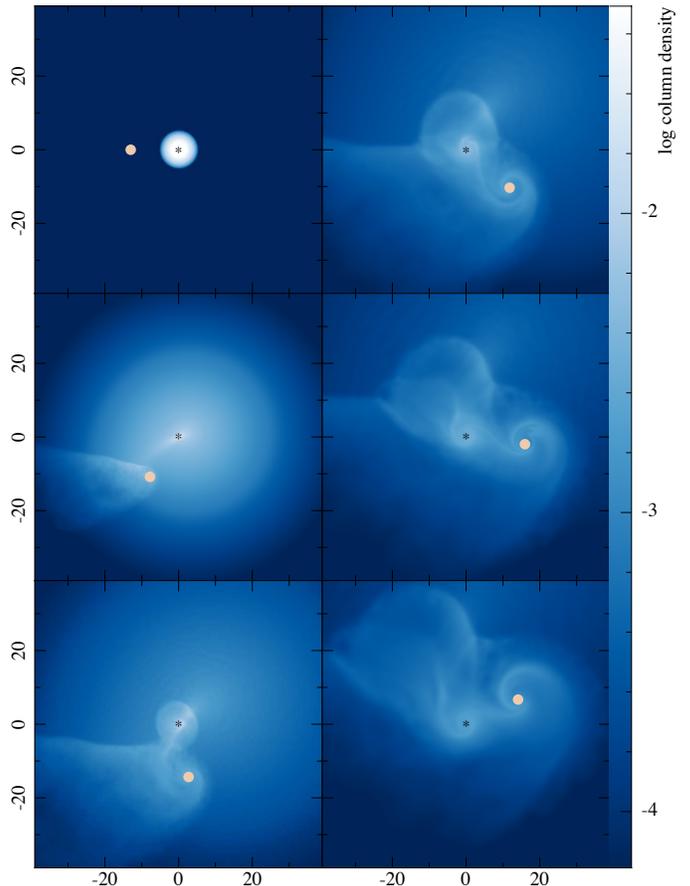}
   \caption{Column density of the  {\it failed} SN ejecta integrated along the binary's rotation axis. The evolution of the companion (white dot) and the ejecta is being followed from the position of the recently formed BH (black dot). The distance scale  is in $R_\odot$ units, and the box covers roughly $40R_{\odot}$ around the BH. The column density is scaled by $\Sigma_{c} = 4.1080\times10^{11}$ g cm$^{-2}$. The panels show the evolution up to $t=1.1$ days, in time steps of $\Delta t = 0.185$ days, increasing from top to bottom and from left to right.}
 \label{fig:Dens}

 \end{center}
 \end{figure}

Figure \ref{fig:Dens} shows the evolution of the integrated column density (along the binary's rotation axis) in the  {\it failed} SN explosion case.  A large amount of material is observed to remain within $\approx 40R_{\odot}$ from the BH (black dot used as the rest frame in all panels) at the end of the simulation  which could be eventually accreted by the BH or the companion. The white dot indicates the position of the $12M_{\odot}$ companion star which  interacts gravitationally with the ejecta and the BH, and is allowed to accrete material using the same criteria that was used for the BH. It can be seen in the last panel from Figure \ref{fig:Dens} that each binary component has  an accretion disk and eventually a circumbinary disk is formed around the binary. By the end of the  fallback phase, the companion accretes $\approx 0.05M_{\odot}$ of low velocity ejecta  coming mostly from the original oxygen layer of the star.  Since explosive nuclear burning occurs within this layer,  the enriched infalling material  will pollute  the companion's atmosphere. Enrichment signatures have already been observed as  overabundances of $\alpha$ elements (O, Mg, Si, S) in the companion stars of X-ray binary systems \citep{Israelian1999, Gonzalez2008,Gonzalez2011, Suarez2015} and might be  explained by the capture of SN ejecta from a close companion.

\begin{figure}
\begin{center}

\includegraphics[width=0.49\textwidth]{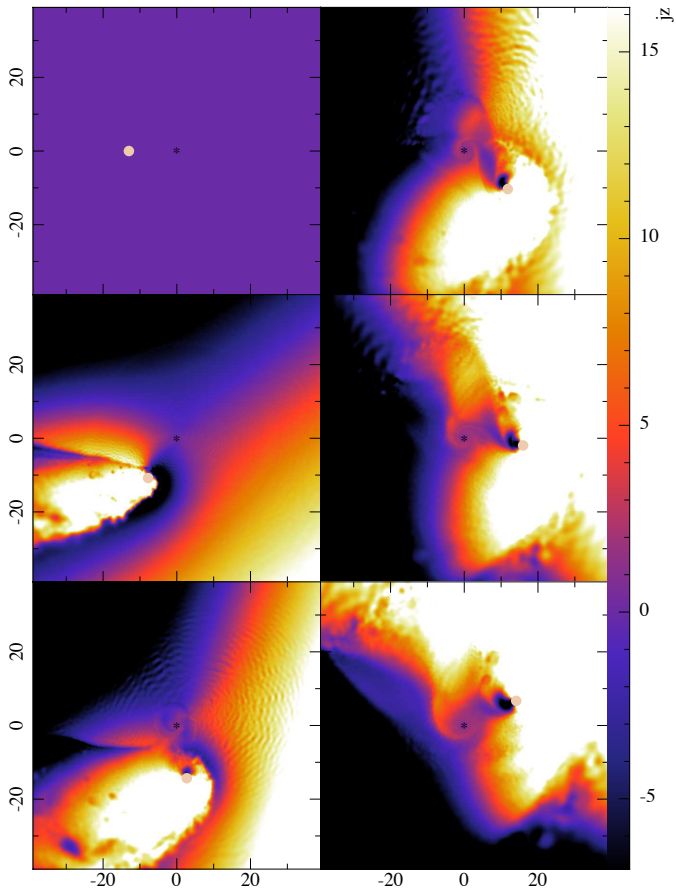}
   \caption{Evolution of the SN ejecta's specific angular momentum within the orbital plane as measured from the position of the BH. The color on the first panel corresponds to an initially small positive angular momentum $j_{\rm z}\gtrsim0$. Blue to black colors indicates negative values of $j_{\rm z}$, while colors ranging from purple to red and white correspond to positive angular momentum larger than the initial value. The specific angular momentum is scaled by $j_{\rm c} = 3.04\times10^{18}$ cm$^{2}$ s$^{-1}$. The panels show the same snapshots as in Figure \ref{fig:Dens}.}
 \label{fig:Jz}

 \end{center}
 \end{figure}

Figure \ref{fig:Jz} shows the specific angular momentum $j_{\rm z}$ of the ejecta measured from the BH's position. Initially all material is contained within $\approx 4R_{\odot}$ and has very small angular momentum. As the ejecta moves away from the recently formed BH, it is torqued by the companion. This causes material to lose or gain angular momentum according to its original position with respect to the companion. Such effect can be seen in the second panel in Figure \ref{fig:Jz} where  two distinct regions are clearly seen; the blue to black regions comprise material with specific angular momentum $j_{\rm z}\leq0$, torqued in the opposite direction of its original angular momentum, while purple to red and white regions comprise material with $j_{\rm z}>0$ which gain angular momentum due to the interaction with the binary. From the third panel on, after about 0.55 days, it is evident that a centrifugally supported structure forms around the BH, whose subsequent accretion is able to spin up the BH.

Both the successful  and  {\it failed} SN explosions, show the formation of an accretion disk around the binary before completing a full orbital period. However, the amount of accreted material  is determined by the velocity profile of the explosion. In order to get a substantial fraction of the ejected mass to form a sizable accretion disk, the velocity profile of the bounded material must have velocities that are comparable to  the escape velocity of the exploding progenitor. Material with $v_{r}\ll v_{\rm esc}$ will not get  far  from the BH and will promptly be accreted, thus drastically reducing  the torque that could get from its interaction with the companion.

\section{Conclusions}
Through a series of 3d SPH simulations, we have studied an scenario in which  a rapidly rotating BH in a HMXRB can be  produced  following a {\it failed}  SN explosion. We have studied three different  scenarios for the SN explosion of the tidally locked stellar progenitor  in a close binary system: direct collapse to a BH, a successful SN  explosion with a large kinetic to binding energy ratio, and  a {\it failed} SN explosion with a low kinetic to binding energy ratio. In the last case,  the resulting  velocity profile  allows for a large fraction of the material to reach distances comparable to the separation of the binary.

Our simulations show that the angular momentum content of the SN material  in a close binary system can be increased through its interaction with the binary companion. Material falling back from beyond the BH's Roche lobe can be significantly torqued by the companion, changing its trajectory and velocity towards the BH and ultimately increasing its  angular momentum.  As shown in the {\it failed} SN explosion case, in order to obtain a large gain in the angular momentum accreted by the BH, it is key for the ejecta  to reach  distances comparable to the binary's separation before falling back, otherwise, the angular momentum gain will be negligible.  We thus conclude that the  presence of rapidly rotating BHs in HMXRBs  can potentially  be explained by invoking  a {\it failed} SN  explosion mechanism, in which a large fraction of the stellar material is marginally bound.

\acknowledgments

\bibliography{BHspin_plus}

\end{document}